\documentclass[11pt]{article}
\usepackage{moriond,epsfig}

\usepackage[]{natbib,amsmath,amssymb}
\usepackage{graphicx}

\newcommand{\aap}[0]{A\&A}

\begin{document}
\vspace*{4cm}
\title{WEIGHTING THE CLUSTERS OF GALAXIES WITH WEAK GRAVITATIONAL LENSING: 
The problem of the mass-sheet degeneracy}
\author{M. Brada\v{c}$^{1,2}$, M. Lombardi$^{1,3}$, P. Schneider$^{1}$} 

\address{$^{1}$Institut f\"ur Astrophysik und Extraterrestrische
  Forschung, Auf dem H\"ugel 71, D-53121 Bonn, Germany
  $^{2}$Max-Planck-Institut f\"ur Radioastronomie, Auf dem H\"ugel 69,
  D-53121 Bonn, Germany  $^{3}$European Southern Observatory,
  Karl-Schwarzschild-Str. 2, D-85748 Garching bei M\"unchen, Germany}

%
%
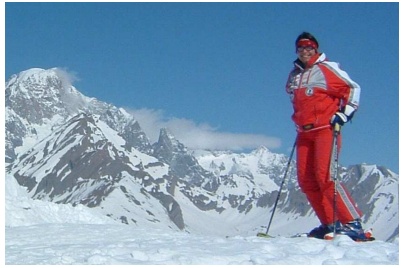
\maketitle

\abstracts{Weak gravitational lensing is considered to be one of the
  most powerful tools to study the mass and the mass distribution of
  galaxy clusters.  However, weak lensing mass reconstructions are
  plagued by the so-called mass-sheet degeneracy~--~the surface mass
  density $\kappa$ of the cluster can be determined only up to a degeneracy
  transformation $\kappa \to \kappa' = \lambda \kappa + (1 -
  \lambda)$, where $\lambda$ is an arbitrary
  constant.  This transformation fundamentally limits the accuracy of
  cluster mass determinations if no further assumptions are made.  We
  discuss here a possibility to break the mass-sheet degeneracy in weak
  lensing mass maps using distortion and redshift information of
  background galaxies.  Compared to other techniques proposed in the
  past, it does not rely on any assumptions on cluster potential and
  does not make use of weakly constrained information (such as the
  source number counts, used in the magnification effect).  Our
  simulations show that \textit{we are effectively able to break the
    mass-sheet degeneracy for supercritical lenses} and 
that for undercritical lenses the mass-sheet degeneracy is very
  difficult to be broken, even under idealised conditions.}

\def\diff{\mathrm{d}}

\def\eck#1{\left\lbrack #1 \right\rbrack}
\def\eckk#1{\bigl[ #1 \bigr]}
\def\rund#1{\left( #1 \right)}
\def\abs#1{\left\vert #1 \right\vert}
\def\wave#1{\left\lbrace #1 \right\rbrace}
\def\ave#1{\left\langle #1 \right\rangle}

\def\vec#1{%
  \if\alpha#1\mathchoice%
    {\mbox{\boldmath$\displaystyle#1$}}%
    {\mbox{\boldmath$\textstyle#1$}}%
    {\mbox{\boldmath$\scriptstyle#1$}}%
    {\mbox{\boldmath$\scriptscriptstyle#1$}}%
  \else
    \textbf{\textit{#1}}%
  \fi}

%
%
\section{Introduction}
\label{sec:introduction}
Weak gravitational lensing techniques have been to great extent
applied to measure the cluster mass distribution.
Unfortunately, all these methods suffer from the fact that the
projected surface mass density $\kappa$ can be determined only up to a
degeneracy transformation $\kappa \to \kappa' = \lambda \kappa + (1-\lambda)$, where $\lambda$ is an
arbitrary constant.  This invariance fundamentally limits the accuracy of cluster mass
determinations if no further assumptions are made.  In particular,
as we will show later this transformation leaves the main observable 
unchanged and 
therefore $\lambda$ can not be
directly constrained.

A naive solution to the problem of mass-sheet degeneracy 
is to constrain $\lambda$ by making simple
assumptions about $\kappa$.  For example, one can assume that the surface mass
density is decreasing with distance from the centre, implying $\lambda >
0$.  In addition, $\kappa$ is likely to be non-negative, and so one can
obtain an upper limit on $\lambda$ (for $\kappa < 1$). 

More quantitatively, with the use of wide field cameras one might try
to assume that $\kappa \simeq 0$ at the boundary of the field, far
away from the cluster center.  However, if we consider for example a
$M_\mathrm{vir} = 10^{15} M_{\odot}$ cluster at redshift $z = 0.2$, we
expect from N-body simulations to have a projected dimensionless
density of about $\kappa \simeq 0.005$ at $15 \mbox{ arcmin}$ from the
cluster center (Douglas Clowe, private communication).  Hence, even
with the use of a $30 \times 30 \mbox{ arcmin}$ camera we expect to
underestimate the virial mass of such a cluster by $\sim 20
\%$. Therefore additional constraints need to be used. We show in
these proceedings (more details can be found in \citep{bradac03prep}),
that background (photometric) redshifts can help us to lift this
degeneracy and therefore remove the fundamental limit on cluster mass
reconstructions.

\section{Principles of weak gravitational lensing}
\label{sc:lens}
Weak gravitational lensing measures the strength of the gravitational
field from a sample of measured ellipticities of background galaxy
images. Under the assumption that the intrinsic ellipticity distribution is
isotropic, $\bigl\langle \epsilon^{\rm s} \bigr\rangle = 0$, the expectation
value for the lensed, image ellipticities at redshift $z$ becomes
\begin{equation}
  \label{eq:7}
  \bigl\langle \epsilon(z) \bigr\rangle = \begin{cases}
    g(\vec\theta, z) & \text{if $\bigl\lvert g(\vec\theta, z) 
      \bigr\rvert < 1 \;,$} \\
    1 / g^*(\vec\theta, z)
    & \text{otherwise$\; .$}
  \end{cases}
\end{equation}
The redshift-dependent
reduced shear $g(\vec\theta, z)$ is given by
\begin{equation}
  \label{eq:4}
  g(\vec\theta, z) = \frac{Z(z) \gamma(\vec\theta)}{1 - Z(z) \kappa(\vec\theta)} \; ,
\end{equation}
where  $Z(z)$ is the so-called ``cosmological weight'' function
\citep{lombardibertin99}. By measuring an ensemble average of the lensed image ellipticities, an
unbiased estimator for the reduced shear can be obtained. The  $Z(z)$  
function accounts for the fact that the sources that have a redshift 
$z$ smaller than the
deflector $z_{\rm d}$ are not lensed ($Z(z \le z_{\rm d})=0$) and 
asymptotically 
increases to 1 for $z \to \infty$.  Note that, as suggested by its name, $Z(z)$
is cosmology dependent.  In \citep{lombardibertin99} 
the authors have shown that the differences
between Einstein-de Sitter and the nowadays assumed standard cosmology
($\Omega_{\rm m} = 0.3$, $\Omega_\Lambda = 0.7$) are not significant for the
purpose of cluster-mass reconstructions.  Therefore we will from now on
use Einstein-de Sitter cosmology.

\subsection{The problem of the mass sheet degeneracy}
\label{sc:mass-sheet}
In the simple case of  background sources all having the {\it same 
redshift}, the
mass-sheet degeneracy can be understood just using the above
equations.  Indeed, consider for a moment 
the transformation of the potential $\psi$
\begin{equation}
  \label{eq:8a}
  \psi(\vec\theta, z) \to \psi'(\vec\theta, z) = 
  0.5\: \rund{1-\lambda} \vec\theta^2 +  \lambda\psi(\vec\theta, z)\; ,
\end{equation}
where $\lambda$ is an arbitrary constant. 
$\kappa$ and $\gamma$ are related to the potential $\psi$ through its
second partial derivatives (denoted by subscript), namely 
$\kappa = 0.5\: (\psi_{,11}+\psi_{,22})$, $\gamma_1 =
  0.5\: (\psi_{,11}-\psi_{,22})$, $\gamma_2 = \psi_{,12}$.
From \eqref{eq:8a} it follows that $\kappa$ transforms as 
\begin{equation}
  \label{eq:8}
  \kappa(\vec\theta, z) \to \kappa'(\vec\theta, z) = 
  \lambda \kappa(\vec\theta, z) + (1 - \lambda) \; ,
\end{equation}
and the shear  $\gamma(\vec\theta, z)$ changes to $\lambda \gamma(\vec\theta, z)$. Therefore
the reduced shear $g(\vec\theta, z)$ (our main observable)  remains invariant.

The authors in \citep{seitz97} have shown that in the case of a 
known {\it redshift distribution}, a similar form of the mass-sheet
degeneracy holds to a very good approximation  for non-critical 
clusters (i.e.\ for clusters
with $|g(\vec \theta, z)| \leq 1$ for all source redshifts).  In
  such a case the standard weak-lensing mass reconstruction is affected by the
degeneracy
\begin{equation}
  \label{eq:9}
  \kappa \to \kappa' \simeq \lambda\kappa+ (1-\lambda) \bigl\langle Z(z) \bigr\rangle / \bigl\langle Z^2(z) \bigr\rangle\; ,
\end{equation}
where $\bigl\langle Z^n(z) \bigr\rangle$ denotes the $n$-th order moment of the
distribution of cosmological weights.  As a result,
\textit{standard\/} weak-lensing reconstructions are still affected by
the mass-sheet degeneracy even for sources at different redshifts;
moreover, simulations show that the degeneracy is hardly broken even
for the lenses close to critical. 

In this work we use the information of {\it individual redshifts} of
background sources
for to break this degeneracy. Indeed, suppose for simplicity that half of the
background sources are located at a known redshift $z^{(1)}$, and the
other half at another known redshift $z^{(2)}$.  Then, the weak lensing
reconstructions based of the two populations will provide two
different mass maps, $\kappa'(\vec\theta, z^{(1)})$ and
$\kappa'(\vec\theta, z^{(2)})$, leading to two different forms of the 
mass-sheet
degeneracy. In other words, the two mass reconstructions $(i=1,2)$ are
given by
\begin{equation}
  \label{eq:901}
  \kappa'(\vec\theta, z^{(i)}) = \lambda^{(i)} \kappa_{\rm
  t}(\vec\theta, z^{(i)}) + \bigl(1 - \lambda^{(i)} \bigr)
\end{equation}
where we have denoted $\kappa_{\rm t}(\vec\theta,z^{(i)})$ 
the true projected $\kappa$
of the lens at the angular position $\vec\theta$ 
for sources at redshift $z^{(i)}$. Since 
the transformation \eqref{eq:901} holds for any
$\vec\theta$, we have  a system of equations
to be solved for $\lambda^{(1)}$ and $\lambda^{(2)}$. 
The relation between $\kappa_{\rm
  t}(\vec\theta,z^{(1)})$ and  $\kappa_{\rm
  t}(\vec\theta,z^{(2)})$ is known, namely $\kappa_{\rm
  t}(\vec\theta,z^{(1)}) Z(z^{(2)}) = \kappa_{\rm t}(\vec\theta,z^{(2)}) Z(z^{(1)})$.
Suppose one  measures
both  $\kappa'(\vec\theta, z^{(i)})$ at $N$ different positions 
$\vec\theta_{j}$, this gives
us a system of  $2N$ equations to be solved for $\lambda^{(i)}$ and 
$\kappa_{\rm t}(\vec\theta_{j})$.  This 
theoretically allows us to break the mass-sheet degeneracy.

It is interesting to observe that
this argument only applies to relatively ``strong'' lenses.  Indeed,
for ``weak'' lenses, i.e.\ lenses for which we can use a first order
approximation in $\kappa$ and $\gamma$, the expectation value of
measured image ellipticities is $\bigl\langle
\epsilon(z) \bigr\rangle = \gamma(\vec\theta, z)$. In such case  
the degeneracy of the form $\psi(\vec\theta, z) \to \psi'(\vec\theta, z) = 
  0.5\: \rund{1-\lambda} \vec\theta^2 +  \psi(\vec\theta, z)$
leaves the observable $\gamma(\vec\theta, z)$ unchanged. 
As a result, the method described above cannot be
used to break the mass-sheet degeneracy for these lenses. Only when
the $(1-Z(z)\kappa)$ term in the reduced shear becomes important and
$g(\vec\theta,z)$ can be distinguished from $\gamma(\vec\theta, z)$ in
the (noisy) data, we will be able to make unbiased cluster mass 
reconstructions.

\section{Results}
\label{sec:simulated-data}
In order to test whether we can break the mass-sheet degeneracy by
using redshift information on the background sources, we performed a
simple test on simulated data.

We assume parametrised model families for the
underlying cluster mass distribution and use the
likelihood analysis (described in detail in \citep{bradac03prep}). 
We consider a non-singular model that approximates an
isothermal sphere for large distances in which we
allow for a constant sheet in surface mass density.  
The dimensionless surface mass density is given by
\begin{equation}
  \label{eq:19}
  \kappa ( \theta / \theta_\mathrm{c} ) = \kappa_0 \rund{1 + \theta^2 /\rund{2 \theta_\mathrm{c}^2}}\rund{1 + \theta^2 / \theta_\mathrm{c}^2}^{-3/2} + \kappa_1 \; ,
\end{equation}
where $\kappa_0$, $\kappa_1$ are dimensionless constants and
$\theta_\mathrm{c}$ is the core radius.

\begin{figure}[ht]
\begin{minipage}{0.5\textwidth}
\begin{center}
\includegraphics[width=0.9\textwidth]{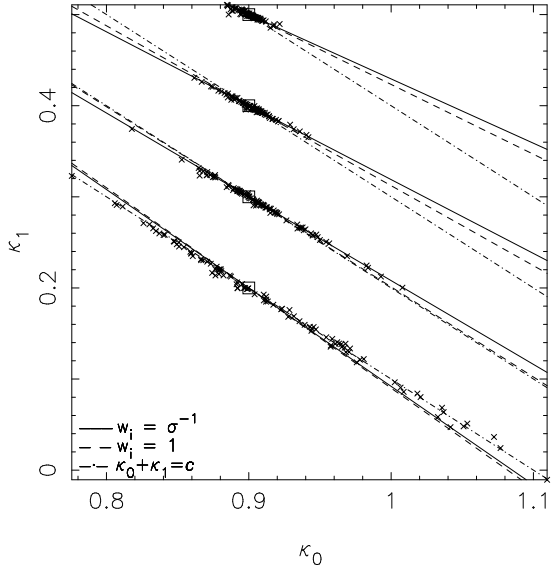}
\end{center}
\end{minipage}
\begin{minipage}{0.5\textwidth}
\vspace{-1cm}
\caption{Recovered parameter values (crosses) as a result of
minimising the log-likelihood function. For each of the
four sets of parameters (denoted by squares) 100 mock catalogues
were created. For these $N_{\rm g} = 2000$ galaxies 
were drawn randomly across
the field of $6 \times 6\mbox{ arcmin}^2$. The intrinsic ellipticities
$\epsilon^\mathrm{s}$ were drawn from a Gaussian  distribution
characterised by $\sigma_{\epsilon^\mathrm{s}} = 0.15$. We draw the redshifts of the background sources following a $\Gamma$-distribution with $z_0 = 2 / 3$. We put the lens at a redshift $z_{\rm d} = 0.2$.
The measurement errors $\epsilon^{\rm err}$ on the observed
ellipticities 
were drawn from a Gaussian distribution with
$\sigma_\mathrm{err} = 0.1$ and added to the lensed ellipticities,
for the redshift errors we use 
$\sigma_\mathrm{zerr} = 0.06 \rund{1+z_i}$. Solid and dashed lines 
correspond to the expected mass-sheet degeneracy calculated using
$w_i= 1 / \sigma_i^2$ and $w_i={\rm const.}$ for the weighting
scheme respectively (both almost overlap). Dot-dashed lines are given by
$\kappa_0 + \kappa_1 ={\rm const}$.}
\label{fig:fitlens}
\end{minipage}
\end{figure}

Figure~\ref{fig:fitlens} shows the results of
log-likelihood minimisation of four different lens parameters, for
each of them 100
mock realisations were calculated.  Solid and dashed lines in the
figure correspond to the expected mass-sheet degeneracy calculated
using Eq.~\eqref{eq:9}. We use different weighting schemes to
calculate $\bigl\langle Z^n(z) \bigr\rangle$. A best fit to the data
is given by weighting each galaxy with the inverse of $\sigma_i^2$
\begin{equation}
  \label{eq:17}
  \sigma_i^2 = \bigl( 1- \bigl\lvert \langle \epsilon \rangle
  \bigr\rvert^2\bigr)^2 \sigma^2_{\epsilon^{\rm s}}
  + \sigma^2_{\rm err}\; ,
\end{equation}
where $\sigma_i$ is an approximation
for the true dispersion of measured ellipticities (see caption of
Fig.~\ref{fig:fitlens} for detailed description).

\section{Conclusions}
\label{sec:conclusions}
We considered a new method to break the mass-sheet
degeneracy in weak lensing mass reconstructions using shape
measurements only. 
Our main conclusions are:\\
(1) The mass-sheet degeneracy can be effectively broken by using
  redshift information of the individual sources.  However, this is
  effective for critical clusters only, i.e.\ for clusters that have
  sizable regions where multiple imaging is possible (and thus perhaps
  observed).  The statistical lensing analysis has to be employed
  close to and inside the critical curves of the cluster.  In the
  regions far outside the critical curve, where weak lensing mass
  reconstructions are normally performed, the lens is too weak for
  the mass-sheet degeneracy to be broken by using redshift and distortion
  information only, even when idealised conditions are employed.\\
(2) Using simulations we find that correlations remaining from the 
mass-sheet degeneracy transformation for
  critical lenses are well described by $\kappa \to \kappa' \simeq (1
  - \lambda) \kappa + (1-\lambda) \bigl\langle Z(z) \bigr\rangle / \bigl\langle Z^2(z) \bigr\rangle$, where the
  moments of the cosmological weights are calculated using
  \eqref{eq:17}. \\
(3) In order to break the mass-sheet degeneracy with current data it
  is necessary to extend the statistical lensing analysis closer to
  the cluster centre and to simultaneously perform weak and strong
  lensing analysis of the cluster.  This will be a subject of a future
  study.

\section*{Acknowledgements}
  This work was
  supported by the IMPRS and BIGS Graduate Schools, by the EU Grant
  for young scientists and by the DFG. 

\bibliographystyle{unsrt}    

\end{document}